\documentclass[prd,reprint,amsmath,nofootinbib,superscriptaddress,showpacs,floatfix]{revtex4-1}

\usepackage[T1]{fontenc}
\usepackage[latin1]{inputenc}
\usepackage[brazil,english]{babel}

\usepackage{amsmath}
\usepackage{graphicx}
\usepackage{amssymb}
\usepackage{esint}
\usepackage[usenames,dvipsnames]{color}

\usepackage{longtable}
\usepackage{dcolumn}

\makeatother

\usepackage{babel}

\makeatletter

\begin{document}

\title{Electromagnetic perturbations in new brane world scenarios}

\author{C. Molina}
\email{cmolina@usp.br}
\affiliation{Escola de Artes, Ci\^{e}ncias e Humanidades, Universidade de S\~{a}o Paulo \\ 
Av. Arlindo Bettio 1000, CEP 03828-000, S\~{a}o Paulo-SP, Brazil}

\author{A. B. Pavan}
\email{alan@unifei.edu.br}
\affiliation{Instituto de F\'{\i}sica e Qu\'{\i}mica, Universidade Federal de Itajub\'a \\ 
Av. BPS 1303 Pinheirinho, 37500-903, Itajub\'a-MG, Brazil}

\author{T. E. Medina Torrej\'on}
\email{temttm@if.usp.br}
\affiliation{Instituto de F\'{\i}sica, Universidade de S\~{a}o Paulo \\
C.P. 66318, 05315-970, S\~{a}o Paulo-SP, Brazil}

\begin{abstract}
In this work we consider electromagnetic dynamics in Randall-Sundrum branes. It is derived a family of four-dimensional spacetimes compatible with Randall-Sundrum brane worlds, focusing on asymptotic flat backgrounds. Maximal extensions of the solutions are constructed and their causal structures are discussed. These spacetimes include singular, non-singular and extreme black holes.
Maxwell's electromagnetic field is introduced and its evolution is studied in an extensive numerical survey. Electromagnetic quasinormal mode spectra are derived and analyzed with time-dependent and high-order WKB methods. Our results indicate that the black holes in the brane are electromagnetically stable.
\end{abstract}



\maketitle

\section{Introduction}

Brane world models have gained attention with the development of string theory. They are meant to be phenomenological implementations of more fundamental stringy models. Although extra-dimensional scenarios already have a long history \cite{Kaluza:1921tu,Klein:1926tv}, the more recent brane perspective of the Universe offers new insights and challenges.

The main characteristics of string inspired brane worlds is that standard model fields are confined to a four-dimensional hypersurface, the brane, while gravity propagates in a larger spacetime, the bulk \cite{lrr-2010-5}. A basic challenge of any brane world model is to explain why gravity, in large scales, is approximately described by standard Newton's law. One approach is to consider that the extra dimensions are compactified. This is the starting point of the Arkani-Hamed-Dimopoulos-Dvali (ADD) models \cite{ArkaniHamed:1998rs}. Nevertheless, brane worlds where the extra dimension is noncompact, can also be implemented. The simplest models in this context are the Randall-Sundrum brane worlds, which describe our Universe as a domain wall embedded in a five dimensional anti-de Sitter spacetime \cite{Randall:1999vf,Randall:1999ee}.

Black holes in standard general relativity and in extra-dimensional scenarios are important sources of gravitational waves, and the recent direct observation of gravitational perturbations \cite{PhysRevLett.116.061102} opened a new window to test extra-dimensional models. Also, according to some setups with extra dimensions, it is possible that small black holes could be produced (and detected) in particles collisions in TeV energy scale \cite{Casanova:2005id,Cavaglia:2002si}, implying that quantum gravity may show itself already at the current particle accelerators.

But, while cosmological brane solutions are abundant, the construction of bulk spacetimes describing compact objects in a brane is not straightforward. For instance, the natural generalization of the Schwarzschild metric to Randall-Sundrum models corresponds to an infinite black string extended through the fifth dimension, whose induced metric on the brane is purely Schwarzschild \cite{Chamblin:1999by}. However, although the curvature scalars are everywhere finite, the Kretschmann scalar diverges at the AdS horizon at infinity, turning the black string into a physically unsuitable object. Also, usual $p$-brane black holes in dimensions higher than four are generically unstable due to the mechanism presented by Gregory and Laflamme \cite{Gregory:1993vy}.


Given the relative complexity in the derivation of exact bulk solutions modeling compact objects in Randall-Sundrum scenarios, one practical alternative is to build four-dimensional geometries in the brane and invoke Campbell-Magaard theorems \cite{Seahra:2003eb}, which guarantees their extensions through the bulk (at least locally).
This approach has been used by several authors in the treatment of compact objects in Randall-Sundrum branes (for example in \cite{Casadio:2001jg,Bronnikov:2003gx,Lobo:2007qi,Molina:2010fb,Molina:2012ay,Dadhich:2000am,Lobo:2007qi,Neves:2015vga}) and will be employed in the present work.

Our description of black holes in a Randall-Sundrum scenario will be done from the point of view of a brane observer. Four-dimensional effective gravitational field equations obtained by Shiromizu, Maeda and Sasaki \cite{Shiromizu:1999wj} are assumed to describe gravity in the brane. Within this context, we derive new spacetimes deforming the vacuum solution with the techniques introduced in \cite{Molina:2010yu,Molina:2011mc,Molina:2012ay,Molina:2013mwa}.
It is interesting to mention that the geometries obtained here could be considered as a subset of the possible solutions that could be algorithmically generated in the approach suggested in \cite{Bronnikov:2003gx} if additional restrictions are imposed. However, following our approach, we have obtained completely integrated exact solutions.

Considering the perturbative aspects of the brane spacetimes, a detailed investigation of scalar and gravitational perturbations in brane world backgrounds was conducted in \cite{Abdalla:2006qj}. In the present work, we extend that previous investigation considering a more phenomenological perturbation, Maxwell's electromagnetic field \cite{Kanti:2005xa,Konoplya:2011qq}, evolving in the new brane geometries derived here.
In particular, we focus on the electromagnetic quasinormal mode spectra since they dominate the decay of the field for intermediate times and can be considered ``footprints'' of specific scenarios. Moreover their measurement could lead to observable signatures discriminating brane models \cite{Seahra:2004fg}.

The structure of this paper follows the points discussed in this introduction.
In Sec.~\ref{spacetimes} we derive a family of analytic asymptotically flat solutions in a Randall-Sundrum brane world. The maximal extensions of the solutions are constructed and analyzed.
The electromagnetic field, together with the methods used to investigate its evolution, are introduced in Sec.~\ref{perturbations}.
The electromagnetic dynamics is numerically treated and the quasinormal spectra are obtained and discussed in Sec.~\ref{dynamics}.
Conclusions are presented in Sec.~\ref{conclusion}, and some convergence issues and limitations of the numerical algorithms are commented in the Appendix.

In this work we have used the metric signature $\textrm{diag}(-+++)$ and the geometric units $G_{4D}=c=1$, where $G_{4D}$ is the effective four-dimensional gravitational constant.

\section{Brane world spacetimes}
\label{spacetimes}

\subsection{Spacetime metric}

In this section we introduce a family of brane world spacetimes, deriving four-dimensional metrics that are compatible with a Randall-Sundrum setup.
Following the approach suggested by Shiromizu, Maeda and Sasaki \cite{Shiromizu:1999wj}, the effective four-dimensional gravitational field equations in a vacuum Randall-Sundrum brane is
\begin{equation}
R_{\mu\nu} - \frac{1}{2} R g_{\mu\nu} = - E_{\mu\nu} \,\, .
\label{eq_projetada}
\end{equation}
We have assumed that the effective four-dimensional cosmological constant in the brane is null. In brane world models, $E_{\mu\nu}$ is proportional to the traceless projection on the brane of the five-dimensional Weyl tensor. Therefore, the four-dimensional Ricci scalar $R$ vanishes,
\begin{equation}
R = 0 \,\, .
\label{Ricci}
\end{equation}

Considering a spherically symmetric and static brane, the metric has the form given by
\begin{equation}
ds^{2} = -A(r)dt^{2} + \frac{dr^{2}}{B(r)} + r^{2} (d\theta^{2} + \sin^{2} \theta d\phi^{2}) \,\, .
\label{metrica_4d}
\end{equation}
In such a case, the vanishing of the Ricci scalar can be written as a constraint between the functions $A$ and $B$,
\begin{equation}
\frac{A''}{A} - \frac{(A')^{2}}{2A^{2}} + \frac{A'B'}{2AB} + \frac{2}{r} \left[\frac{A'}{A} + \frac{B'}{B}\right]
= \frac{2(1-B)}{r^{2} B} \,\, ,
\label{constraint}
\end{equation}
with prime ($'$) denoting differentiation with respect to $r$.

The most general solution for the metric which solves Eq.~\eqref{constraint} with the condition $A=B$ is
\begin{equation}
A_{0}(r) = B_{0}(r) = 1 - \frac{2M}{r} + \frac{q}{r^{2}}  \,\, .
\label{vac_solution}
\end{equation}
The metric defined by the functions $A=A_{0}$ and $B=B_{0}$ in Eq.~\eqref{vac_solution} has the same form of the Reissner-Nordstr\"{o}m metric. However, in this case $q$ is a ``tidal charge'' related to the structure of the brane and not to an electric charge. In addition, there is no restriction to the sign of $q$, though there is an upper limit for $q$. The constant $M$ is the mass of the black hole, and since we are interested in solutions which describe black holes, it is assumed that $q < M^{2}$. From now on, this metric will be called our base solution.

We propose to construct solutions ``close'' to those represented by the Eq.~(\ref{vac_solution}). More precisely, we seek a family of spacetimes which are one-parameter continuous deformations of the considered base geometry. In addition, we require that the set of new solutions includes the base solution. For this aim, we write
\begin{equation}
A(r) = A_{0}(r) \,\, ,
\label{ansatz_A}
\end{equation}
\begin{equation}
B(r) = B_{0}(r) + (C - 1) \, B_{lin}(r) \,\, ,
\label{ansatz_B}
\end{equation}
where $C$ is a constant.

The vanishing of the four-dimensional Ricci scalar $R$ can be interpreted as a linear constraint involving the components of $E_{\mu\nu}$ so that the solution for the correction $B_{lin}$ can be obtained with the same techniques presented in \cite{Molina:2013mwa}. Up to an arbitrary multiplicative constant, $B_{lin}$ is given by
\begin{equation}
B_{lin}(r) = \exp \left[
- \int \frac{f(r)}{A_{0}(r) h(r)} \, dr
\right] \,\, ,
\label{Blin}
\end{equation}
with the functions $f$ and $h$ given by
\begin{eqnarray}
f(r) & = & r A_{0}(r) A''_{0}(r) - \frac{r \left[ A_{0}(r) \right]^{2}}{2}
+ 2 A_{0}(r) A'_{0}(r) + 2 \frac{ A_{0}^{2}(r)}{r}
\nonumber \\
& = & \frac{2}{r^{5}}
\left[
r^{4} - 4Mr^{3} + 3(M^{2}+q)r^{2} - 4Mqr + q^{2}
\right] \,\,,
\end{eqnarray}
and
\begin{equation}
h(r) = \frac{r A'_{0}(r)}{2} + 2 A_{0}(r)
 = 2 - \frac{3M}{r} + \frac{q}{r^{2}} \,\, .
\end{equation}

The roots of the function $A_{0}$ are relevant, being given by
\begin{equation}
r_{+} = M + \sqrt{M^{2} - q} \,\,\, , \,\,\,
r_{-} = M - \sqrt{M^{2} - q} \,\, .
\label{rp}
\end{equation}
Also relevant are the roots of the function $h$,
\begin{eqnarray}
r_{0} & = & \frac{1}{4}\left[3M + \sqrt{9M^{2}-8q}\right] \,\, , 
\nonumber \\
r_{0-} & = & \frac{1}{4}\left[3M - \sqrt{9M^{2}-8q}\right] \,\, .
\label{r0}
\end{eqnarray}
With the constants $r_{+}$, $r_{-}$, $r_{0}$ and $r_{0-}$, the complete solutions for $A$ and $B$ can be expressed as
\begin{equation}
A(r) = \frac{(r-r_{+})(r-r_{-})}{r^{2}}\,\,,
\label{sol_A}
\end{equation}
\begin{eqnarray}
B(r) & = & \frac{(r - r_{+})(r - r_{-})}{r^{2}} \nonumber \\
& & \times \left[ 1 +
\left( C - 1 \right) \,
\frac{(r_{+} - r_{0})^{c_{0}} \, (r_{+} - r_{0-})^{c_{0-}}}{(r - r_{0})^{c_{0}}
\, (r - r_{0-})^{c_{0-}}} \right] \,\, ,
\label{sol_B}
\end{eqnarray}
where the coefficients $c_{0}$ and $c_{0-}$ are given by
\begin{equation}
c_{0} = \frac{1}{2} + \frac{3}{2\sqrt{9-8\frac{q}{M^{2}}}} \,\,\, , \,\,\,
c_{0-} = \frac{1}{2} - \frac{3}{2\sqrt{9-8\frac{q}{M^{2}}}} \,\, .
\label{c0m}
\end{equation}

The geometries defined by Eqs.~\eqref{sol_A} and \eqref{sol_B} include, as particular cases, some solutions already presented in the literature. For instance, the CFM metric \cite{Casadio:2001jg,Bronnikov:2003gx} is obtained from Eqs.~\eqref{sol_A} and \eqref{sol_B} setting $M>0$ and $q=0$. Also, the ``zero mass black hole'' metric \cite{Bronnikov:2003gx} is recovered with $M=0$ and $q<0$.

\subsection{Global structure}

Once the explicit solution for the metric is obtained, the next step is to characterize the global properties of the spacetimes considered. This is done in the present section.

Considering the limit $r \rightarrow \infty$ of the metric functions $A$ and $B$ in Eqs.~\eqref{sol_A} and \eqref{sol_B}, we have
\begin{equation}
A(r) \sim B(r) \sim 1 + o \left( \frac{1}{r} \right) \,\, .
\end{equation}
Hence, the associated spacetimes are asymptotically flat for any value of $C$. In spite of this fact, the causal structure of the geometry is highly dependent of the this parameter.

For instance, if $C>1$ then $A(r)>0$ and $B(r)>0$ when $r>r_{+}$, and the coordinate system $(t,r,\theta,\phi)$ is valid in this region. The analytic extension beyond $r=r_{+}$ can be made, for example, with the ingoing and outgoing Eddington charts $(u,t,\theta,\phi)$ and $(v,t,\theta,\phi)$, where $u$ and $v$ are the light-cone variables $u = t - r_{\star}$ and $v = t + r_{\star}$.
The radial variable $r_{\star}$ is the so-called ``tortoise coordinate'', defined as
\begin{equation}
\frac{dr_{\star}(r)}{dr} = \frac{1}{\sqrt{A(r)\: B(r)}} \,\, .
\label{tortoise_coordinate}
\end{equation}
The surface $r=r_{+}$ is a Killing horizon and also an event horizon, and therefore the spacetime constructed describes a black hole. The surface gravity $\kappa_{+}$ associated with the Killing horizon is given by
\begin{equation}
\kappa_{+} =
\frac{\sqrt{M^{2} - q}}{\left( M + \sqrt{M^{2} - q} \right)^{2}} \, \sqrt{C} \,\, .
\label{surface_gravity}
\end{equation}

An important feature of the brane geometry with \linebreak $C>1$ comes from the fact that since $c_{0}>0$ the function $B$ is divergent at $r_{0}$, and
\begin{equation}
0 < r_{0} < r_{+} \,\, ,
\label{desigualdade}
\end{equation}
as it can be directly verified from Eqs.~\eqref{rp} and \eqref{r0}.
In the maximal extension, the metric is well defined in the region $r_{0} < r < r_{+}$, but the curvature invariants are not bounded, as seen by the behavior of the Kretschmann scalar near $r_{0}$:
\begin{equation}
\lim_{r\rightarrow r_{0}} \left| R_{\alpha\beta\gamma\delta}R^{\alpha\beta\gamma\delta} \right| \rightarrow \infty \,\, .
\end{equation}
Therefore, for this case the curvature singularity is located at $r\rightarrow r_{0}$. Due to the inequality in Eq.~\eqref{desigualdade}, the interior region of the black hole is singular, but the singularity is always hidden by the event horizon.

\begin{figure}[h]
\includegraphics[clip,width=\columnwidth]{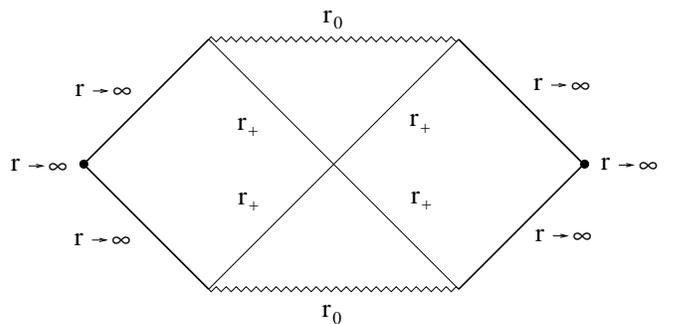}
\caption{Penrose diagram for the brane world spacetimes describing singular black holes.}
\label{diagrama_Cmaior1}
\end{figure}

Using standard techniques (see, for example, \cite{Walker:1970}), Penrose diagrams can be constructed.
For brane world spacetimes with $C>1$, describing singular black holes, the associated Penrose diagrams have the form presented in Fig.~\ref{diagrama_Cmaior1}.

If $C=1$ we recover the base solution. The causal structure of this spacetime is well known (see for example \cite{hawking1975large}), being similar to the case previously discussed, with $C>1$.

If $0 < C < 1$, we still have a black hole surrounded by an event horizon, with an associated surface gravity $\kappa_{+}$ given in Eq.~\eqref{surface_gravity}. But the black hole interior has a more complex structure, when compared to the regime where $C \ge 1$. The function $B$ is not positive-definite between $r_{0}$ and $r_{+}$, having a simple zero at $r=r_{min}$, with
\begin{equation}
0 < r_{0} < r_{min} < r_{+} \,\, .
\end{equation}

\begin{figure}[h]
\includegraphics[clip,width=\columnwidth]{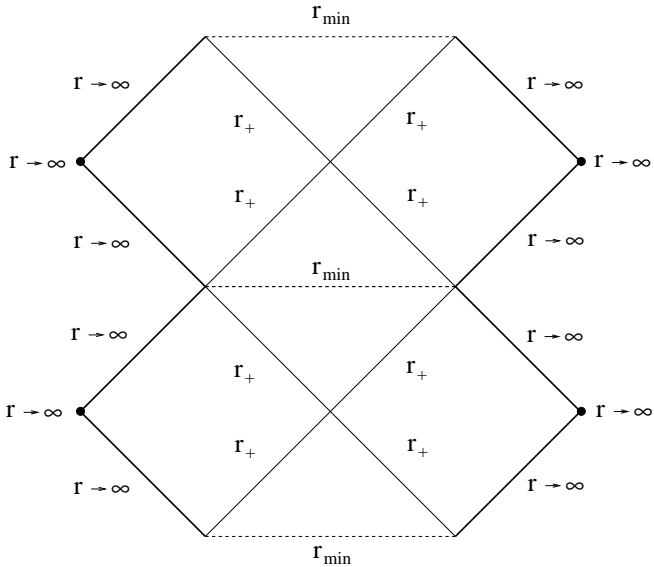}
\caption{Penrose diagram for the brane world spacetimes describing regular black holes.}
\label{diagrama_Cmenor1}
\end{figure}

\noindent
The analytic extension beyond $r=r_{min}$ is suggested with the use of the proper length $L$ as radial function, where
\begin{equation}
\frac{dL(r)}{dr} = \frac{1}{\sqrt{B(r)}} \,\, .
\label{proper_length}
\end{equation}
Choosing an appropriate integration constant in Eq.~\eqref{proper_length}, the region $r_{min}<r<r_{+}$ is mapped into $0<L<L_{max}$, with a finite $L_{max}$. An analytic extension is defined continuing the metric with $-L_{max}<L<L_{max}$.
Since $B$ is bounded in $r_{min} < r< r_{+}$, the spacetime is regular everywhere.
We have a black hole with a regular interior, with no singularity present. The Penrose diagram for the regular black hole spacetime ($0<C<1$) is presented in Fig.~\ref{diagrama_Cmenor1}.

\begin{figure}[t]
\includegraphics[clip,width=0.9\columnwidth]{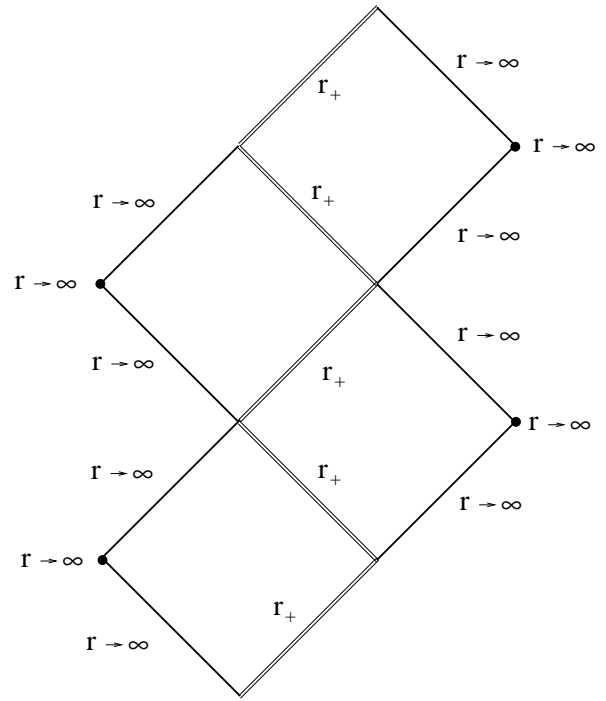}
\caption{Penrose diagram for the brane world spacetimes describing extreme black holes.}
\label{diagrama_Cigual0}
\end{figure}

If $C=0$, the Killing horizon $r=r_{+}$ becomes extreme. The derived solution models a black hole with a null surface gravity. This is possible even with $q<M^{2}$ (the usual non-extreme condition for the base solution). The extension beyond $r=r_{+}$ in this case can be made, for example, using the ``quasi-global'' radial coordinate $w$ \cite{Bronnikov:2003gx,Bronnikov:2008by}, defined as
\begin{equation}
\frac{dw(r)}{dr} = \sqrt{\frac{A(r)}{B(r)}} \,\, .
\label{quasi_global}
\end{equation}
The Penrose diagram for the extreme black hole spacetime ($C=0$), is presented in Fig.~\ref{diagrama_Cigual0}.

\begin{figure}[b]
\includegraphics[clip,width=0.7\columnwidth]{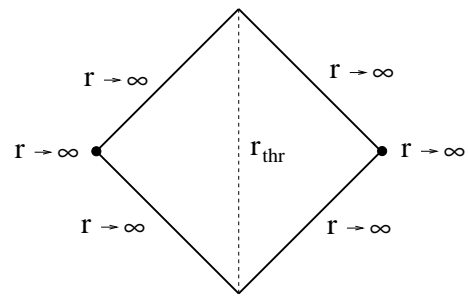}
\caption{Penrose diagram for the brane world spacetimes describing wormholes.}
\label{diagrama_Cmenor0}
\end{figure}

If $C<0$, the function $B$ has two simple positive zeros, $r_{+}$ and $r_{thr}$, where
\begin{equation}
0 < r_{+} < r_{thr} \,\, ,
\end{equation}
and $A(r_{thr})\ne0$. A possible choice of radial coordinate to perform the extension beyond $r=r_{thr}$ is the proper length $L$, in a similar way to what was done when $0<C<1$.
\noindent
The zero $r_{+}$ of the functions $A$ and $B$ does not play any role in the interior causal structure since $r_{+}<r_{thr}$. No event horizon is present and the curvature invariants are finite in everywhere in the spacetime.
In the maximal extension, the surface $r=r_{thr}$ is a time-like outer trapping horizon, so we can conclude that the geometry has a wormhole structure.
The wormhole throat $r=r_{thr}$ connects to asymptotically flat regions and its associated Penrose diagram is presented in Fig.~\ref{diagrama_Cmenor0}.

Considering the Penrose diagrams shown in Figs.~\ref{diagrama_Cmaior1}-\ref{diagrama_Cmenor0}, we see that the general behaviour of the causal structures described in this work were already anticipated in \cite{Bronnikov:2003gx}, although in that analysis the authors did not consider specifically the case $A = A_{0}$ in Eq.~\eqref{vac_solution}. This is no surprise, since the new spacetimes presented here are a subset of the possible solutions that could be generated with the algorithms developed in \cite{Bronnikov:2003gx}.

\section{Electromagnetic perturbations}
\label{perturbations}

\subsection{Axial and polar components of the electromagnetic field}

With the spacetime determined, the next step is to investigate its response to small perturbations. In lowest order, background backreaction can be disregarded, and dynamics is restricted to the matter perturbations in a fixed geometry.
In the present work we will consider a perturbation with direct phenomenological interest, namely, Maxwell's electromagnetic field.

The temporal evolution of a sourceless electromagnetic field in a curved spacetime, minimally coupled to the geometry, is driven by Maxwell's equations,
\begin{equation}
\nabla_{\nu} F^{\nu\mu} = 0 \,\, .
\label{maxequation}
\end{equation}
From the classical electromagnetic tensor $F_{\mu\nu}$, the potential $A_{\mu}$ is defined as
\begin{equation}
F_{\mu\nu} = \partial_{\mu} A_{\nu} - \partial_{\nu} A_{\mu} \,\, .
\label{maxtensor}
\end{equation}

We follow the approach used in \cite{Ruffini:1972pw}, but it is important to stress that some of our expressions will not coincide with the ones presented in that reference. On the other hand, our results are consistent with \cite{PhysRevD.63.124008,Baldiotti:2014pca}, where the issue was treated from a different perspective aiming quantum field theory applications.
In order to make the discrepancies clear, and also to fix notation, we will review the approach in \cite{Ruffini:1972pw} and present our development in some detail.

In spherically symmetric spacetimes, the electromagnetic potential $A_{\mu}$ can be expanded in vector spherical harmonics as
\begin{eqnarray}
A_{\mu}(t,r,\theta,\phi) = \sum_{\ell,m}
\left(\left[\begin{array}{c}
0\\
0 \\
\frac{a(r,t)}{\sin\theta}\ \frac{\partial Y^{\ell m}}{\partial \phi}\\
-a(r,t)\ \sin\theta\ \frac{\partial Y^{\ell m}}{\partial \theta}
\end{array}\right] \right. \nonumber \\
+
\left. \left[\begin{array}{c}
f(r,t) Y^{\ell m} \\
h(r,t) Y^{\ell m} \\
k(r,t)\ \frac{\partial Y^{\ell m}}{\partial \theta} \\
k(r,t)\ \frac{\partial Y^{\ell m}}{\partial \phi}
\end{array}\right]
\right) \,\, ,
\label{decomp}
\end{eqnarray}
where $Y^{\ell m}$ are the scalar spherical harmonics.
The first and second columns in Eq.~\eqref{decomp} are the axial and polar components of the proposed expansion \cite{Ruffini:1972pw}, with parities $(-1)^{\ell+1}$ and $(-1)^{\ell}$, respectively.

Rewriting Maxwell's equations~\eqref{maxequation}, using the metric functions $A$ and $B$ defined in Eq.~\eqref{metrica_4d}, we obtain
\begin{gather}
\partial_{t} F^{\mu t} + \sqrt{\frac{B(r)}{A(r)}} \frac{1}{r^2}
\partial_{r} \left( r^2\sqrt{\frac{A(r)}{B(r)}} F^{\mu r} \right) 
\nonumber \\
+ \frac{1}{\sin \theta} \partial_{\theta} (\sin\theta F^{\mu \theta})
+\partial_{\phi} F^{\mu \phi} = 0 \,\, .
\label{eqF}
\end{gather}
Combining Eqs.~\eqref{maxtensor}-\eqref{eqF}, two different sets of equations will be obtained for the electromagnetic field. These two groups describe the axial and polar modes, which will be treated separately in the following.

\subsubsection{Axial electromagnetic modes}

The non-zero axial mode components of the electromagnetic tensor $F^{\mu\nu}$ are%
\footnote{The component $F^{ \theta \phi}$ presented in \cite{Ruffini:1972pw} has a misprint that was corrected here.}
\begin{eqnarray}
 F^{t\theta} & = & \frac{-1}{A(r)r^2 \sin \theta}\partial_t a^{\ell m}\partial_\phi Y^{\ell m} \,\, ,
\label{1F-1l+1} \\
F^{t\phi} & = & \dfrac{1}{A(r)r^2 \sin \theta} \partial_t a^{\ell m} \partial_\theta Y^{\ell m} \,\, ,
\label{2F-1l+1} \\
F^{r\theta} & = & \frac{ B(r)}{r^2 \sin \theta}\partial_r a^{\ell m} \partial_\phi Y^{\ell m} \,\, ,
\label{3F-1l+1} \\
F^{r\phi} & = & -\dfrac{B(r)}{r^2 \sin \theta} \partial_r a^{\ell m} \partial_\theta Y^{\ell m} \,\, ,
\label{4F-1l+1} \\
F^{ \theta \phi} & = & \dfrac{\ell(\ell+1)}{r^4 \sin \theta} a^{\ell m} Y^{\ell m} \,\, .
\label{5F-1l+1}
\end{eqnarray}
Combining results \eqref{1F-1l+1}-\eqref{5F-1l+1} with Eq.~\eqref{eqF}, we obtain a single relation,
\begin{eqnarray}
& -& \frac{\partial^2 a^{\ell m}}{\partial t^2}
+\sqrt{A(r)B(r)}\frac{\partial}{\partial r}\left[\sqrt{A(r)B(r)}\frac{\partial}{\partial r} a^{\ell m}\right] \nonumber \\
& & = \frac{\ell(\ell+1)}{r^2} a^{\ell m}A(r) \,\, .
\label{alpha}
\end{eqnarray}
The partial differential equation~\eqref{alpha} determines the evolution of the axial electromagnetic perturbation. The master variable in this case is $a^{\ell m}$, and this relation agrees with one presented in \cite{Ruffini:1972pw}.

Axial equation of motion~\eqref{alpha} can be written in a simpler form if the tortoise radial coordinate, introduced in Eq.~\eqref{tortoise_coordinate}, is used. Redefining the master variable as
\begin{equation}
\Psi_{axial}(t,r_{\star}) = a^{\ell m} (t,r(r_{\star})) \,\, ,
\end{equation}
Eq.~\eqref{alpha} takes the form
\begin{equation}
-\frac{\partial^{2} \Psi_{axial}}{\partial t^2} + \frac{\partial^{2} \Psi_{axial}}{\partial r_{\star}^2}  = V_{axial} \, \Psi_{axial}  \,\, ,
\end{equation}
where the axial effective potential $V_{axial}$ is given by
\begin{equation}
V_{axial} (r_{\star}) =  A(r(r_{\star})) \frac{\ell(\ell+1)}{\left[ r(r_{\star}) \right]^{2}} \,\, .
\label{pot_axial}
\end{equation}

\subsubsection{Polar electromagnetic modes}

The nonzero components of $F^{\mu\nu}$ for the polar electromagnetic perturbations are%
\footnote{We found that the components $F^{t\phi}$ and $F^{r\phi}$ presented in \cite{Ruffini:1972pw} have misprints in their denominators that were corrected here.}
\begin{eqnarray}
F^{tr} & = & \frac{B(r)}{A(r)}(\partial_r f^{\ell m}-\partial_t h^{\ell m})Y^{\ell m} \,\, ,
\label{1F-1l} \\
F^{t\theta} & = & -\frac{1}{A(r)r^2}(\partial_t k^{\ell m}-f^{\ell m}) \partial_\theta {Y^{\ell m}} \,\, ,
\label{2F-1l} \\
F^{t\phi} & = & -\frac{1}{A(r)r^2 \sin ^2 \theta} (\partial_t k^{\ell m}-f^{\ell m})\partial_\phi { Y^{\ell m}} \,\, ,
\label{3F-1l} \\
F^{r\theta} & = & \frac{B(r)}{r^2}(\partial_r k^{\ell m}-h^{\ell m}) \partial_\theta {Y^{\ell m}} \,\, ,
\label{4F-1l} \\
F^{r\phi} & = & \dfrac{B(r)}{r^2 \sin ^2 \theta} (\partial_r k^{\ell m}-h^{\ell m}) \partial_\phi Y^{\ell m} \,\, .
\label{5F-1l}
\end{eqnarray}
Substituting Eqs.~\eqref{1F-1l}-\eqref{5F-1l} into Eq.~\eqref{eqF}, three independent coupled equations are derived,
\begin{gather}
\sqrt{A(r)B(r)}\partial_r\left[r^2\sqrt{\frac{B(r)}{A(r)}}(\partial_r f^{\ell m}-\partial_t h^{\ell m})\right] \nonumber \\
-\ell (\ell+1)(\partial_t k^{\ell m}-h^{lm}) =0 \,\, ,
\label{10a1}
\end{gather}
\begin{equation}
\frac{1}{A(r)}\partial_t(\partial_t h^{\ell m}-\partial_r f^{\ell m})- \frac{\ell(\ell+1)}{r^2}(\partial_r k^{\ell m}-h^{\ell m})
= 0 \,\, ,
\label{10b1}
\end{equation}
\begin{gather}
\sqrt{\frac{B(r)}{A(r)}}\partial_{r}\left[\sqrt{A(r)B(r)}(h^{\ell m}-\partial_r k^{\ell m})\right] \nonumber \\
- \frac{1}{A(r)}\partial_t(f^{\ell m}-\partial_t k^{lm})
=0 \,\, .
\label{10c1}
\end{gather}

The next step is to decouple Eqs.~\eqref{10a1}-\eqref{10c1}, producing one single master equation for the polar perturbation. For this purpose, a master variable $b^{\ell m}$ (proposed in \cite{Ruffini:1972pw}) is defined as
\begin{equation}
\partial_t h^{\ell m}-\partial_r f^{\ell m} =
\frac{\ell(\ell+1)}{r^2}b^{\ell m} \,\, .
\label{12}
\end{equation}
Substituting Eq.~\eqref{12} into Eq.~\eqref{10a1} we obtain
\begin{equation}
\sqrt{A(r)B(r)}\partial_r\left(\sqrt{\frac{B(r)}{A(r)}}b^{\ell m}\right)-(\partial_t k^{\ell m}-f^{\ell m})=0 \,\, .
\label{12a}
\end{equation}
Also, combining Eq.~\eqref{12} and Eq.~\eqref{10b1},
\begin{equation}
-\frac{1}{A(r)} \partial_t b^{\ell m}-(h^{\ell m}-\partial_r k^{\ell m}) = 0 \,\, .
\label{12b}
\end{equation}
Finally, differentiating Eq.~\eqref{12a} with respect to $r$, differentiating Eq.~\eqref{12b} with respect to $t$, summing up the results and using Eq.~\eqref{12}, we get
\begin{gather}
 A(r) \partial_{r}\left[\sqrt{A(r)B(r)}\partial_r\left(\sqrt{\frac{B(r)}{A(r)}}b^{\ell m}\right)\right]
- \partial^{2}_{t} b^{\ell m} \nonumber \\
= A(r) \frac{\ell(\ell+1)}{r^2}b^{\ell m} \,\, .
\label{b}
\end{gather}
This is our differential equation for the evolution of the polar perturbation, in terms of the master variable $b^{\ell m}$.
This result does not agree with the one found in \cite{Ruffini:1972pw} when $g_{tt} \neq g_{rr}^{-1}$, that is, when $A \ne B$. But, as we will see, the result presented here agrees with the development found in \cite{PhysRevD.63.124008}.

We now write the polar equation of motion in terms of the tortoise coordinate. Using Eq.~\eqref{tortoise_coordinate} we have
\begin{gather}
\frac{1}{\sqrt{A(r)B(r)}} \frac{\partial^2}{\partial r^2_{\star}}\left(\sqrt{\frac{B(r)}{A(r)}}\widetilde{b}^{\ell m}\right)
-\frac{1}{A(r)} \frac{\partial^2 \widetilde{b}^{\ell m} }{\partial t^2}
\nonumber \\
= \frac{\ell(\ell+1)}{r^2}\widetilde{b}^{\ell m} \,\, ,
\label{c}
\end{gather}
where $\widetilde{b}^{\ell m}(t,r_{\star})={b}^{\ell m}(t,r(r_{\star}))$.
We define a new master variable $\Psi_{polar}$ as
\begin{equation}
\Psi_{polar}(t,r_{\star}) = \sqrt{\frac{B(r)}{A(r)}} \, \widetilde{b}^{\ell m}(t,r(r_{\star})) \,\, .
\end{equation}
Eq.~\eqref{c} is then written as
\begin{equation}
-\frac{\partial^2 \Psi_{polar}}{\partial t^2}
+\frac{\partial^2 \Psi_{polar}}{\partial r^2_{\star}}
=V_{polar}\Psi_{polar} \,\, ,
\label{eq_onda_a2}
\end{equation}
with the polar effective potential $V_{polar}$ given by
\begin{equation}
V_{polar} (r_{\star}) = A(r(r_{\star})) \frac{\ell(\ell+1)}{\left[ r(r_{\star}) \right]^{2}} \,\, .
\label{pot_polar}
\end{equation}

In summary, employing the approach introduced in \cite{Ruffini:1972pw}, we have demonstrated that both axial and polar modes of the electromagnetic perturbation can be described by the same effective potential, as indicated in Eqs.~\eqref{pot_axial} and \eqref{pot_polar}. Moreover, the function $B$ does not explicitly appear in the expression for the effective potentials, even when $A \ne B$ ($g_{tt} \ne g_{rr}^{-1}$).
Because of this property, one could think that the bulk effects on the brane do not affect the electromagnetic field dynamics because both effective potentials (axial and polar) apparently do not depend of the function $B$. This is not true since both metric functions $A$ and $B$ are used in the definition of $r_{\star}$ and therefore influence the characteristics of $V_{axial}(r_{\star})$ and $V_{polar}(r_{\star})$.

\subsection{Numerical methods}

\subsubsection{Direct integration of the Cauchy problem}

The fact that the equations of motion for the axial and polar perturbations can be written in the same form, with the same effective potential, greatly simplifies the treatment of the electromagnetic perturbation.
It follows that the analysis is the same for axial and polar components of the field. In order to simplify notation, we denote by $\Psi$ and $V$ both the axial and polar master variables ($\Psi_{axial}$, $\Psi_{polar}$) and effective potentials ($V_{axial}$, $V_{polar}$). In fact, we effectively have only one equation of motion to consider:
\begin{equation}
- \frac{\partial^{2} \Psi}{\partial t^{2}}
+ \frac{\partial^{2} \Psi}{\partial r^{2}_{\star}}
= V(r_{\star}) \Psi \,\, .
\label{eq_diff}
\end{equation}

The hyperbolic equation~\eqref{eq_diff} can be treated as a Cauchy problem. In this formulation, initial data are given by two functions $F$ and $G$, where
\begin{equation}
\Psi \left(0,r_{\star}\right) = F\left(r_{\star}\right)\,\,,
\end{equation}
\begin{equation}
\frac{\partial\Psi}{\partial t}\left(0,r_{\star}\right)
= G\left(r_{\star}\right)\,\,.
\end{equation}
Since we are interested in the black hole response to localized perturbations, we will consider initial conditions with a sharp peak and fast decay.
For the most of numerical evaluations presented here, the initial data have the form
\begin{equation}
F\left(r_{\star}\right) = A_{1} \, e^{-\sigma_{1} r_{\star}^{2}} \,\,\, , \,\,\,
G\left(r_{\star}\right) = A_{2} \, e^{-\sigma_{2} r_{\star}^{2}} \,\, .
\end{equation}
From results on black hole oscillations in general relativity and some of its extensions \cite{Nollert:1999ji,Kokkotas:1999bd,Abdalla:2005hu,Abdalla:2006qj,Molina:2010fb,Konoplya:2011qq}, we expect that the main characteristics of the time-evolution profiles, after a transient initial regime, are insensitive to the choice of the initial data, provided that the initial conditions are localized.
In order to check if this actually occurs in the present case, and rule out any eventual influence of initial data on late time results, we have considered different choices for the initial data.

We employ an explicit finite difference scheme to numerically integrate the field equation~\eqref{eq_diff}. A discretized version of Eq.~\eqref{eq_diff} is obtained with
\begin{equation}
t\rightarrow t_{i}=t_{0}+i\,\Delta t \,\,\, , \,\,\, i=0,1,2,\ldots   \,\,,
\label{descreteT}
\end{equation}
\begin{equation}
r_{\star}\rightarrow x_{j}=x_{0}+j\,\Delta x \,\,\, , \,\,\, j=0,1,2,\ldots   \,\,.
\label{descreteX}
\end{equation}
With this discretization, the differential equation is approximated by
\begin{equation}
\psi_{N} = \left( 2 - \Delta t^{2}\, V_{C}\right) \,
\psi_{C} - \psi_{S} + \frac{\Delta t^{2}}{\Delta x^{2}}
\left(\psi_{E} - 2\psi_{C}+\psi_{W}\right) \,\, ,
\label{psiN}
\end{equation}
where
\begin{gather}
\Psi_{N} = \Psi\left(t_{i+1},x_{j}\right) \,\,\, , \,\,\,
\Psi_{E} = \Psi\left(t_{i},x_{j+1}\right) \,\, , \nonumber \\
\Psi_{C} = \Psi\left(t_{i},x_{j}\right) \,\,\, , \,\,\,
\Psi_{W} = \Psi\left(t_{i},x_{j-1}\right) \,\, , \nonumber \\
\Psi_{S} = \Psi\left(t_{i-1},x_{j}\right) \,\,\, , \,\,\,
V_{C} = V_{\ell} \left(t_{i},x_{j}\right) \,\, .
\end{gather}

In the integration method employed, a given section of the plane $t-r_{\star}$ is replaced by a discretized version. This integration grid is illustrated in Fig.~\ref{plano_discreto}. Using the initial conditions (functions $F$ and $G$), the first two lines of the grid are determined. Then, with the discrete wave equation~\eqref{psiN}, the electromagnetic perturbation is integrated. Additional comments about the numerical integration are presented in the Appendix.

\subsubsection{Quasinormal frequencies and WKB approach}

Of particular interest in the perturbative dynamics are the quasinormal spectra. Let us consider a wave function $\Psi(t,r_{\star})$, in the present case the axial or polar electromagnetic perturbation.
The frequency domain wave function $\psi(r_{\star})$ is obtained by a Laplace transform of the function $\Psi$ as
\begin{equation}
\psi(r_{\star}) = \int_{0}^{\infty} \Psi(t,r_{\star}) \, e^{i\omega t}\, dt \,\, ,
\end{equation}
with $\omega$ extended to the complex plane. The time-independent version of Eq.~\eqref{eq_diff} is given by
\begin{equation}
\frac{\partial^{2} \psi}{\partial r_{\star}^{2}} + \left(\omega^{2} - V\right) \psi = 0 \,\, .
\label{wave_equation_frequency}
\end{equation}
Quasinormal modes are solutions of Eq.~\eqref{wave_equation_frequency} satisfying both ingoing and outgoing boundary conditions:
\begin{equation}
\lim_{r_{\star}\rightarrow\mp\infty} \psi \, e^{\pm i\omega r_{\star}} = 1 \,\, .
\end{equation}

\begin{figure}[h]
\includegraphics[clip,width=\columnwidth]{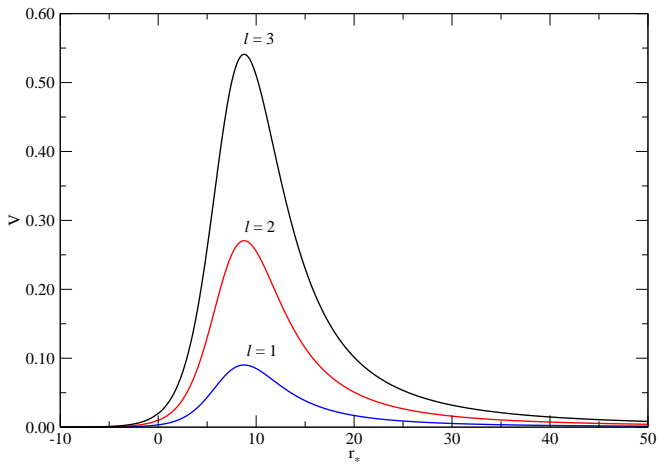}
\caption{Typical profiles for the electromagnetic effective potential. The parameters used were $M=1$, $q=0.5$, $C=1.5$, and $\ell=1,2,3$.}
\label{fig_pot}
\end{figure}

In the present work, we calculate the  quasinormal frequencies directly from the numerical integration of the field equations. Also, we use a high-order WKB approach \cite{Schutz:1985km,Iyer:1986np,Konoplya:2003ii} to compute the frequencies, which have been applied in a variety of situations.
This time-independent algorithm is very efficient when the effective potential has the form of a potential barrier with a single maximum, which asymptotically decays to zero.
These are the main characteristics of the electromagnetic potential defined in Eqs.~\eqref{pot_axial} and \eqref{pot_polar}, as illustrated in Fig.~\ref{fig_pot}.

The WKB high-order formula for the frequencies $\omega$ is given by
\begin{equation}
i \frac{\omega^2 - V_{0}}{\sqrt{-2V_{0}^{\prime\prime}}} - L_{2} - L_{3} - L_{4} -
L_{5} - L_{6} = n + \frac{1}{2} \, ,
\end{equation}
where $V_{0}$ is the maximum value of the potential and $V_0^{\prime\prime}$ is the second derivative of the potential with respect to the tortoise coordinate $r_{\star}$, calculated at the maximum of $V$. Constants $L_{2}$, $L_{3}$  $L_{4}$, $L_{5}$ and $L_{6}$ are presented in \cite{Schutz:1985km,Iyer:1986np,Konoplya:2003ii}. The overtone number is indicated by $n$, with fundamental quasinormal frequencies labeled by $n=0$.

A note of caution should be added when using this high-order WKB approach. It is, \textit{a priori}, difficult to guarantee the convergence of the method. Here we circumvent this difficulty validating the WKB results with those that were extracted in the direct integration evaluations.
As seen in the following section, the WKB formulas converge up to a certain limit value of $C$, for a fixed $\ell$. In this case, the concordance of the direct integration and WKB results is very good. But for high enough $C$, the WKB approach does not appear to converge. More on this issue is discussed in the Appendix.

\section{Perturbative dynamics and stability analysis}
\label{dynamics}

\subsection{Overview of the results}

The analytic and numerical results presented in Sec.~\ref{perturbations} were used in an extensive investigation of the general characteristics of the perturbative electromagnetic dynamics around the spacetimes derived in Sec.~\ref{spacetimes}.

The main point observed is the fact the perturbations are stable. That is, considering localized initial conditions, the function $\Psi(t,r_{\star})$ is always bounded. In particular, for the quasinormal frequencies, we have always that $\textrm{Im}(\omega) < 0$.
Stability follows from the fact that the electromagnetic effective potentials in Eqs.~\eqref{pot_axial} and \eqref{pot_polar} are positive-definite.
This result is corroborated by our numerical investigation.

\begin{figure}[t]
\includegraphics[clip,width=\linewidth]{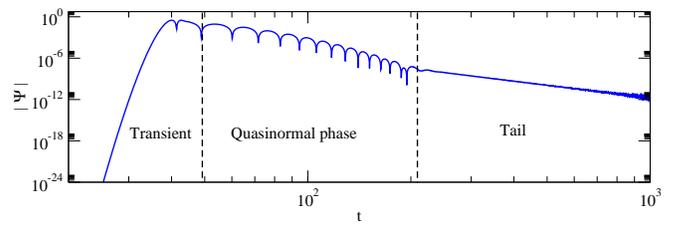}
\caption{Main features of the electromagnetic dynamics. Transient, quasinormal mode and tail phases are pointed out. The parameters used in this particular plot were $\ell=1$, $C=1.5$, $M=1$, and $q=0.5$, but the qualitative aspects presented here are generic.}
\label{overview}
\end{figure}

The qualitative picture observed for the perturbative dynamics is consistent with the one presented in \cite{Abdalla:2006qj} for scalar and gravitational perturbations.
Considering the wave function as seen by an static observer located a fixed value of its radial coordinate ($r_{\star} = r_{\star}^{fixed}$), the first phase is transient, depending on the details of the initial conditions. The transient regime is followed by a quasinormal mode dominant phase. Finally, for late times, the decay is dominated by a power law tail. These considerations are illustrated in Fig.~\ref{overview}.

\subsection{Dependence of the quasinormal frequencies with $C$}

Scanning the parameter space of the model, we analyzed the dependence of the fundamental quasinormal frequency with the parameter $C$.
Specific frequencies for several values of $C$ and $\ell$ are displayed in Table~\ref{qnm_1}, using both the direct integration and 6th-order WKB approaches. The concordance of the methods is very good when the WKB converge.

It should be mentioned that the values presented were obtained with a careful convergence analysis. Details on convergence issues for the methods employed are presented in the Appendix.

\begin{figure}[hb]
\includegraphics[clip,width=\linewidth]{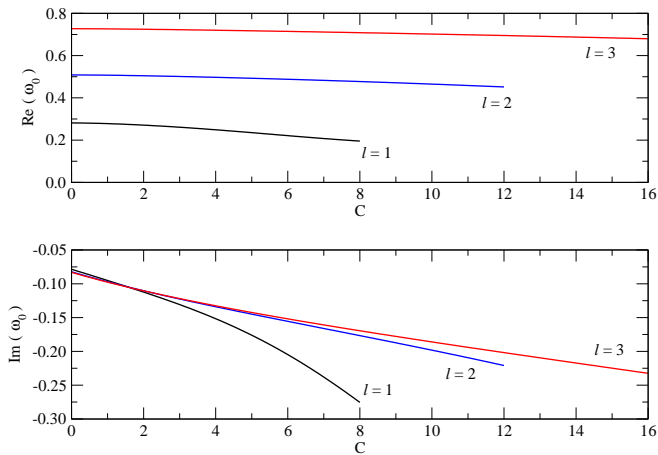}
\caption{Real and imaginary components of the fundamental quasinormal frequencies for several values of $C$. For the considered spacetimes, $\ell=1,2,3$, $M=1.0$, and $q=0.5$.}
\label{w-c}
\end{figure}

In our numerical results, we observed that for a fixed value of $\ell$ the dependence of the fundamental frequency real part is mild in the considered range of $C$. Typically the magnitude of $\textrm{Re}(\omega_{0})$ decays as $C$ increases, that is, the period of oscillation increases with the parameter $C$.
On the other hand, the variation of the imaginary part of fundamental frequency is more robust, with the magnitude of $\textrm{Im}(\omega_{0})$ increasing with $C$.
Therefore, typically electromagnetic perturbations decay faster in spacetimes with higher $C$. This fact could make the detection of quasinormal modes associated to deformed solutions with $C>1$ harder, when compared to the usual general relativity vacuum solution. But geometries with $0\le C < 1$ should be more accessible.
In Fig.~\ref{w-c} we illustrate the points presently considered.

\begin{table}[t]
\caption{Fundamental quasinormal frequencies for the electromagnetic perturbation with several values of $C$ and $\ell$, calculated with both the direct integration approach and WKB formulas. For the considered spacetimes, $M=1.0$ and $q=0.5$.}
\label{qnm_1}
\begin{tabular*}{\columnwidth}{*{6}{c@{\extracolsep{\fill}}}}
\hline
\multicolumn{2}{l}{}                           &
\multicolumn{2}{c}{Direct Integration}         &
\multicolumn{2}{c}{WKB-$6th$ order}                \\ \\
$\ell$ & $C$ & $\textrm{Re} \left( \omega_{0} \right)$ & $\textrm{Im} \left( \omega_{0}\right)$ & $\textrm{Re} \left( \omega_{0} \right)$ & $\textrm{Im} \left( \omega_{0}\right)$ \\
\hline
1 & 0.05 & 0.2809 & -0.08107  & 0.2813 & -0.07941 \\
1 & 0.1  & 0.2804 & -0.08092  & 0.2813 & -0.08022 \\
1 & 0.5  & 0.2803 & -0.08718  & 0.2803 & -0.08677 \\
1 & 1.0  & 0.2784 & -0.09484  & 0.2780 & -0.09513 \\
1 & 1.5  & 0.2744 & -0.1019   & 0.2748 & -0.1036 \\
1 & 5.0  & 0.2411 & -0.1302   & 0.2353 & -0.1761 \\ \\
2 & 0.05 & 0.5092 & -0.08341  & 0.5085 & -0.08303 \\
2 & 0.1  & 0.5079 & -0.08398  & 0.5084 & -0.08320 \\
2 & 0.5  & 0.5078 & -0.08981  & 0.5081 & -0.08985 \\
2 & 1.0  & 0.5073 & -0.09671  & 0.5072 & -0.09704 \\
2 & 1.5  & 0.5061 & -0.1038   & 0.5060 & -0.1038 \\
2 & 5.0  & 0.4912 & -0.1411   & 0.4927 & -0.1449 \\
2 & 10.0 & 0.4672 & -0.1750   & 0.4651 & -0.1982 \\
\hline
\end{tabular*}
\end{table}

\subsection{Dependence of the quasinormal frequencies with $q$}

Continuing our analysis of the electromagnetic perturbative dynamics, we analyze the dependence of the fundamental quasinormal frequencies with the parameter $q$.

Specific values of the fundamental quasinormal frequency $\omega_{o}$ for several values of $q$ and $\ell$ are presented in Table~\ref{qnm_2}, obtained with the direct integration and 6th-order WKB methods. These values were obtained following the convergence analysis presented in the Appendix.

\begin{table}[ht]
\caption{Fundamental quasinormal frequencies for the electromagnetic perturbation with several values of $q$ and $\ell$, calculated with both the direct integration approach and WKB formulas. For the considered spacetimes, $M=1.0$ and $C=1.5$.}
\label{qnm_2}
\begin{tabular*}{\columnwidth}{*{6}{c@{\extracolsep{\fill}}}}
\hline
\multicolumn{2}{l}{}                           &
\multicolumn{2}{c}{Direct Integration}         &
\multicolumn{2}{c}{WKB-$6th$ order}                \\ \\
$\ell$ & $q$ & $\textrm{Re} \left( \omega_{0} \right)$ & $\textrm{Im} \left( \omega_{0}\right)$ & $\textrm{Re} \left( \omega_{0} \right)$ & $\textrm{Im} \left( \omega_{0}\right)$ \\
\hline
1 & -10.0 & 0.1216 & -0.06412   & 0.1164 & -0.06711 \\
1 & -5.0  & 0.1501 & -0.07720   & 0.1452 & -0.07925 \\
1 & -1.0  & 0.2068 & -0.09337   & 0.2045 & -0.09718 \\
1 & -0.5  & 0.2217 & -0.09762   & 0.2204 & -0.1002  \\
1 & 0.5   & 0.2744 & -0.1019    & 0.2748 & -0.1036  \\
1 & 0.9   & 0.3195 & -0.09415   & 0.3198 & -0.09421 \\
1 & 0.95  & 0.3275 & -0.09025   & 0.3274 & -0.09033 \\ \\
2 & -10.0 & 0.2402 & -0.06983   & 0.2399 & -0.07004 \\
2 & -5.0  & 0.2934 & -0.08128   & 0.2931 & -0.08153 \\
2 & -1.0  & 0.3954 & -0.09781   & 0.3953 & -0.09798 \\
2 & -0.5  & 0.4213 & -0.1005    & 0.4212 & -0.1007  \\
2 & 0.5   & 0.5061 & -0.1038    & 0.5060 & -0.1038  \\
2 & 0.9   & 0.5748 & -0.09579   & 0.5749 & -0.09582 \\
2 & 0.95  & 0.5875 & -0.09248   & 0.5875 & -0.09237 \\
\hline
\end{tabular*}
\end{table}

\begin{figure}[hb]
\includegraphics[clip,width=\linewidth]{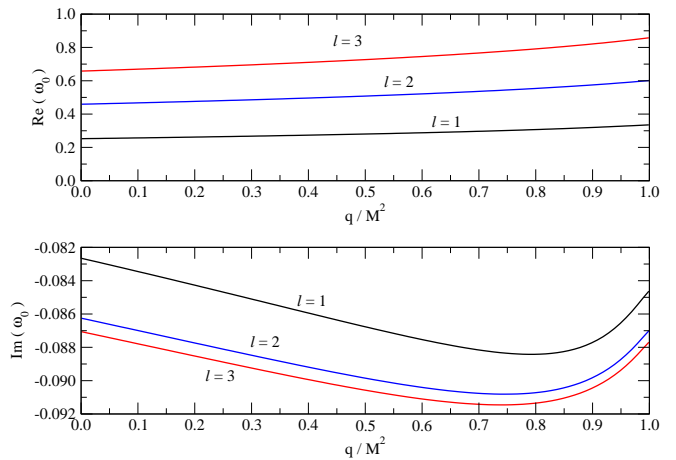}
\caption{Real and imaginary components of the fundamental quasinormal frequencies for several values of $q$. For the considered spacetimes, $\ell=1$, $M=1.0$, and $C=1.5$.}
\label{w-q}
\end{figure}

We have scanned the whole range of $q/M^{2}$, $0<q/M^{2}<1$ for fixed values of $\ell$ and $C$. The qualitative aspect of the dependence of the fundamental frequency is well illustrated in Fig.~\ref{w-q}.
We observe that, while $\textrm{Re}(\omega_{o})$ monotonically grows with $q/M^{2}$, the dependence of $\textrm{Im}(\omega_{o})$ is more complex, typically with a local extremum for some value of  $q/M^{2}$.

\subsection{Beyond the fundamental mode}

We also consider the dependence of the quasinormal frequencies with the overtone number $n$.
For $n>0$, the direct integration usually is not practical. In this case, we rely only on the WKB method. Some care must be taken to ensure that the WKB formulas converge, as discussed the Appendix. As expected, we see that convergence improves as larger values of $\ell$ are considered.

\begin{figure}[t]
\includegraphics[clip,width=\linewidth]{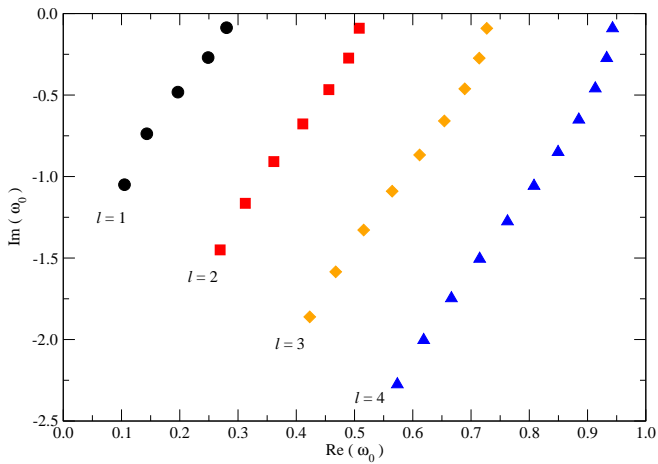}
\caption{Fundamental quasinormal frequencies and overtones for several values of $\ell$ and $M=1.0$, $C=1.5$, and $q=0.5$. For each value of $\ell$, the overtone number increases from top to bottom.}
\label{RewImw}
\end{figure}

A general feature of the results obtained is that $\textrm{Re}(\omega_{n})$ decreases as $n$ increases. That is, the period of oscillation is a monotonically crescent function of $n$, as far as it can be determined with the WKB approach.
Also, the modes are labeled so that the absolute value of $\textrm{Im}(\omega_{n})$ increases with $n$.
These general features are illustrated in Fig.~\ref{RewImw}.

\section{Final comments}
\label{conclusion}

We consider four-dimensional spacetimes compatible with Randall-Sundrum brane world models.
The approach used here to find new black hole solutions was formalized in \cite{Molina:2010yu,Molina:2011mc,Molina:2012ay,Molina:2013mwa}, and it can be seen as complementary to the black hole construction algorithms presented in \cite{Casadio:2001jg,Bronnikov:2003gx}. 
However, in the present development, the problem is treated from a different perspective when compared with the previously cited works. We are interested in spacetimes that can be close to the usual vacuum geometries. Hence, they should be continuous deformations of the vacuum solution. This condition is directly employed, allowing exact and completely integrated solutions to be obtained. 
The generated spacetimes describe wormholes and singular, regular and extreme black holes.

Considering the response of the background to perturbations, the present work continues the development of \cite{Abdalla:2006qj}, where scalar and gravitational disturbances were addressed. Presently, we treat a perturbation with direct phenomenological interest, namely, Maxwell's electromagnetic field.
In our approach to the description of the electromagnetic field, we employ a strategy previously used by the authors of \cite{Ruffini:1972pw} and developed by many others in spherically symmetric geometries where $g_{tt} = g_{rr}^{-1}$. We remark that some of our results do not coincide with \cite{Ruffini:1972pw} when $g_{tt} \ne g_{rr}^{-1}$ (in our notation, $A \ne B$). On the other hand, after an appropriate change in the perturbative master variable, our final expressions are compatible with the ones presented in \cite{PhysRevD.63.124008}.

An important result here is that the spacetimes considered are stable considering this perturbation.
In fact, the qualitative description presented in this work is consistent with the overall picture discussed by the authors in \cite{Abdalla:2006qj}.

After an extensive numerical analysis using a finite difference scheme for the direct integration of the field equation, complemented by a semianalytic high-order WKB method, a general description of some main aspects of the electromagnetic dynamics is presented. Quasinormal spectra are calculated and discussed in detail.
The numeric and semianalytic approaches are complementary, having distinct conditions of applicability. In a wide region of the parameter space (taking the constant $C$ up to a limiting value) we observed a good concordance between both approaches. We consider this result a strong argument for the validation of both methods.

However, for very large values of $C$, the WKB approach does not appear to converge, as the order of the method is increased. This is an interesting point, since the method is widely used in the pertinent literature. A conclusion that can be taken is that although the WKB method is simple and usually reliable, its use without a complementary independent method can be risky.
We stress that a conservative view was adopted in the numerical development in this work and are presented only results where both the direct integration and WKB methods clearly converge and show good agreement.

\begin{acknowledgments}

C. M. is supported by Conselho Nacional de Desenvolvimento Cient\'{\i}fico e Tecnol\'{o}gico (CNPq-Brazil), grant 303431/2012-1. A. B. P. is supported by Conselho Nacional de Desenvolvimento Cient\'{\i}fico e Tecnol\'{o}gico (CNPq-Brazil), grant 472660/2013-6.

\end{acknowledgments}

\appendix

\section*{\textbf{APPENDIX:} Considerations on the convergence of the methods}
\label{appendix}

An important element for the validation of a numerical procedure is a convergence analysis.
In this appendix we discuss some convergence issues associated to both the direct integration routine and the high-order WKB method used.

\begin{figure}[t]

\includegraphics[width=0.9\columnwidth]{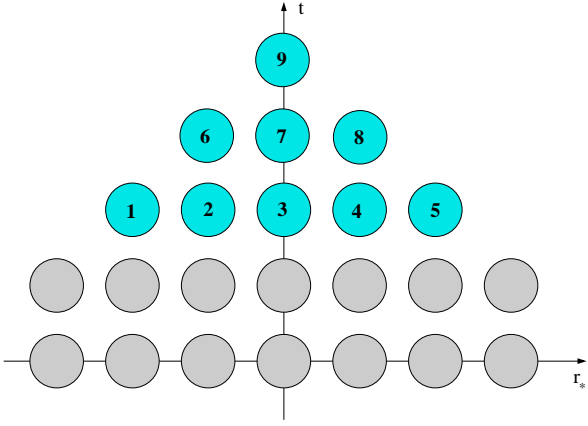}

\caption{Integration grid for the $t-r_{\star}$ plane. In this diagram, $N_{x}=5$.The non-numbered circles indicate the grid points with the initial conditions. The numbers indicate a possible order in which the integration is performed on the grid.}
\label{plano_discreto}

\end{figure}

The integration algorithm was performed with an explicit, finite-difference method, using a discretized version of the $t-r_{\star}$ plane. The integration grid, determined by Eqs.~\eqref{descreteT} and \eqref{descreteX}, is illustrated in Fig.~\ref{plano_discreto}.

\begin{figure}[h]
\includegraphics[clip,width=\linewidth]{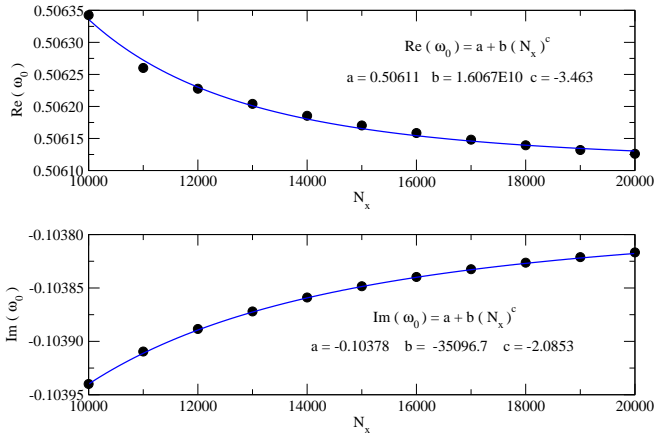}
\caption{(Top) Convergence of the real component of the fundamental quasinormal frequency $\omega_{0}$ with the increase of $N_{x}$. (Bottom) Convergence of the imaginary component of the fundamental quasinormal frequency $\omega_{0}$ with the increase of $N_{x}$. In both graphs, dots represent the integration data and continuous lines represent a fit in the form $a + b\,(N_{x})^{c}$. The parameters for the graphs were $M=1.0$, $q=0.5$, $C=1.5$, and $\ell=2$.}
\label{converge}
\end{figure}

The main parameters describing the integration grid are $N_{x}$ (number of points in the grid base), $r_{\star}^{max}$ (the interval for the variable $r_{\star}$ in the grid base assumes values in the interval $[-r_{\star}^{max},+r_{\star}^{max}]$), and $\Delta t$ and $\Delta r_{\star}$ in Eqs.~\eqref{descreteT} and \eqref{descreteX}.
Typical values used were:

\begin{eqnarray}
N_{x} & & = \textrm{from } 10000 \textrm{ to } 20000 \nonumber\\
r_{\star}^{max} & & =  2000 \nonumber \\
\Delta t / \Delta r_{\star} & & = 0.4
\end{eqnarray}

In the present work, we use as grid size parameter the number of points in the grid base, $N_{x}$. The total points in the grid is approximately $\left( N_{x} \right)^{2}/2$.
Inspecting the results, we have observed that the fundamental frequencies converge with the increase of $N_{x}$ as
\begin{equation}
\omega = a + b\,\left( N_{x} \right)^{c} \,\,,
\end{equation}
with $c<0$, for large enough $N_{x}$.

The determination of the constants $a$, $b$ and $c$ gives us a method improve truncation error limitations, extrapolating (to a certain extent) the finite grid size results to the continuum,
\begin{equation}
\omega \rightarrow  a \,\,\, \textrm{with} \,\,\, N_{x} \rightarrow \infty \,\,.
\end{equation}
Typical convergence curves are presented in Fig.~\ref{converge}.

\begin{figure}[h]
\includegraphics[clip,width=\linewidth]{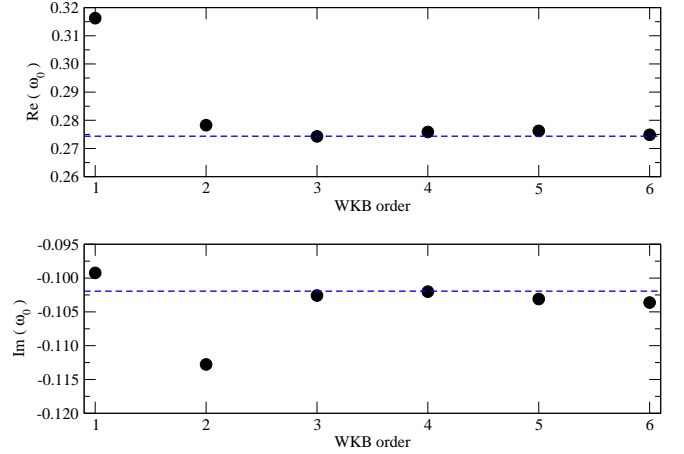}
\caption{Convergence of the real and imaginary components of the fundamental quasinormal frequency $\omega_{0}$ with the WKB order. The doted lines represent the direct integration results. The parameters used are $M=1$, $q=0.5$, $C=1.5$, and $\ell=1$.}
\label{WKBconverge}
\end{figure}

We comment now on the WKB approach used in the present work. This algorithm is very efficient when the potential $V(r_{\star})$ has an isolated maximum and decays sufficiently fast to zero as $r_{\star} \rightarrow \pm\infty$. Such conditions are satisfied by the electromagnetic effective potentials derived in Sec.~\ref{perturbations}.

On the other hand, it is not possible \textit{a priori} to guarantee that the algorithm converge as the order of the method is increased. In the scenario considered here, we have verified that the WKB formulas appear not to approach a definite limit for high enough values of $C$.
We illustrate a typical case where it is observed convergence in the WKB method in Fig.~\ref{WKBconverge}.

We stress that when we observe convergence in the WKB approach, the concordance of the direct integration and WKB results are very good. We consider this result a strong argument for the validation of both approaches. This point is verified in Tables~\ref{qnm_1} and \ref{qnm_2} for instance.
But for values of $C$ where the WKB frequencies do not tend to a well-defined limit, the comparison of this method with the direct integration is poor.
In the present work, we have adopted the conservative view of present only results where the direct integration and WKB methods clearly converge and show good agreement.


%

\end{document}